\documentclass[conference,a4paper]{IEEEtran}

\usepackage{balance}

\ifCLASSINFOpdf
  \usepackage[pdftex]{graphicx}
\else
  \usepackage[dvips]{graphicx}
\fi
\usepackage{amsmath}
\usepackage{siunitx}

\ifCLASSOPTIONcompsoc
 \usepackage[caption=false,font=normalsize,labelfont=sf,textfont=sf]{subfig}
\else
 \usepackage[caption=false,font=footnotesize]{subfig}
\fi

\usepackage{multirow}
\usepackage{textcomp}
\usepackage[dvipsnames]{xcolor}
\usepackage{tikz}
\usepackage{circuitikz}
\usepackage{pgfplots}
\pgfplotsset{compat=1.8}
\usetikzlibrary{decorations.markings}
\usetikzlibrary{shapes.geometric}

\newcommand\copyrighttext{%
\footnotesize \textcopyright~2026 IEEE. Personal use of this material is permitted.  Permission from IEEE must be obtained for all other uses, in any current or future media, including reprinting/republishing this material for advertising or promotional purposes, creating new collective works, for resale or redistribution to servers or lists, or reuse of any copyrighted component of this work in other works.}

\newcommand\copyrightnotice{%
\begin{tikzpicture}[remember picture,overlay]
\node[anchor=south,yshift=10pt] at (current page.south) {\fbox{\parbox{\dimexpr0.75\textwidth-\fboxsep-\fboxrule\relax}{\copyrighttext}}};
\end{tikzpicture}%
}

\begin{document}
\title{A Scalable Reconfigurable Intelligent Surface with 3~Bit Phase Resolution and High Bandwidth for 3.6~GHz 5G/6G Applications}

\author{\IEEEauthorblockN{
Markus Heinrichs\IEEEauthorrefmark{1},
Aydin Sezgin\IEEEauthorrefmark{2},
Rainer Kronberger\IEEEauthorrefmark{1}
}

\IEEEauthorblockA{\IEEEauthorrefmark{1}% 1st affiliations
(TH Cologne): High Frequency Laboratory, TH Köln - University of Applied Sciences, Cologne, Germany,\\markus.heinrichs@th-koeln.de, rainer.kronberger@th-koeln.de}
\IEEEauthorblockA{\IEEEauthorrefmark{2}% 2nd affiliations
(Ruhr University Bochum): Ruhr Universität Bochum, Bochum, Germany, aydin.sezgin@rub.de}
}

\maketitle

\begin{abstract}
Reconfigurable Intelligent Surfaces enable active control of wireless propagation channels, which is crucial for future 5G and 6G networks. This work presents a scalable RIS design operating at 3.6~GHz with both 1~bit and 3~bit phase resolution, supporting wideband applications. The unit cells employ low-cost printed circuit board technology with an innovative spring-contact feeding structure, enabling efficient assembly and reduced manufacturing complexity for large-area arrays. The design achieves broadband phase control, low power consumption, and high scalability, with experimental results demonstrating phase tunability across the n78 frequency band and competitive reflection performance compared to existing solutions. This RIS architecture provides a practical platform for experimental studies of smart radio environments, beam steering, and sensing applications in next-generation wireless networks.
\copyrightnotice
\end{abstract}

\vskip0.5\baselineskip
\begin{IEEEkeywords}
reconfigurable intelligent surface, RIS, intelligent reflecting surface, IRS, 6G.
\end{IEEEkeywords}

\section{Introduction}
% no \IEEEPARstart
In 6G networks, controllable radio environments—often referred to as smart radio environments (SREs)—are considered one of the key innovations. The goal is to actively influence the environment in which radio channels are formed, rather than merely adapting to it. Reconfigurable Intelligent Surfaces (RIS) or Intelligent Reflecting Surfaces (IRS) provide the means to achieve this by allowing the control of wireless propagation channels. Such surfaces can be used for beam steering, joint communication and sensing (JCS) applications, or to improve the security of communication networks \cite{DiRenzo}. For the development of such applications, RIS hardware is required to validate, test, and improve the proposed concepts under real conditions.

This work presents an RIS design for the n78 frequency band at 3.6 GHz, which can be scaled to large-area arrays covering several square meters. This enables experimental studies on how RIS structures affect radio channels and propagation environments in real-world settings. Additionally, special attention was given to minimizing power consumption to ensure usability in large-scale array deployments. The proposed RIS further features a broadband phase response, supporting scientific investigations in wideband 5G and 6G scenarios.

In the following section, the proposed unit cell design is introduced, comprising the antenna element and the phase-switching circuits with \SI{1}{\bit} and \SI{3}{\bit} resolution. The implementation of the realized arrays is described in Section~\ref{sec:realization}, followed by the measurement in Section~\ref{sec:measurements}. The proposed RIS is compared to other designs in Section~\ref{sec:comparison} before the main findings of this work are summarized in Section~\ref{sec:conclusion}.

\section{Unit Cell Design} \label{sec:unitCell}
An RIS consists of an array of identical unit cells, each providing independent control of the reflection phase. Each unit cell comprises an antenna element designed to receive the incident free-space wave that is impinging on the surface and to convert it into a conducted wave, which is coupled to a microstrip line located on the backside of the unit cell. A circuitry providing an electronically switchable reflection coefficient is connected to the microstrip line port of the antenna. This causes the incident wave to be reflected with a phase shift that depends on the switching state of the circuitry. The reflected wave is subsequently re-radiated by the antenna element. Two types of phase-switching circuits have been developed: one with a 1-bit phase resolution, enabling a \SI{180}{\degree} phase change, and another with a 3-bit resolution, allowing phase control in \SI{45}{\degree} steps.

\subsection{Antenna Element}
A linearly polarized rectangular patch antenna was selected as the radiating element. Patch antennas can be fabricated using standard printed circuit board (PCB) technology, which enables cost-effective realization in large quantities. As the bandwidth of a patch antenna can be increased by selecting a larger distance between the patch and the ground plane, and to minimize dielectric losses within the substrate, the unit cell’s antenna is realized using two separate PCBs that are spaced apart by an air gap. The first PCB carries the patch element, while the second PCB serves as the antenna’s ground plane. To further reduce manufacturing costs, low-cost standard FR4 material has been chosen as the substrate, avoiding the need for more expensive RF laminates. The dissipation factor of the FR4 material used is $\tan \delta = 0.025$, which is higher than that of commonly used RF materials such as RO4003C ($\tan \delta = 0.0027$) or RO3003 ($\tan \delta = 0.0010$) from manufacturer Rogers. However, since most of the volume between the patch and the ground plane consists of air, the cost savings achieved by using FR4 outweigh the increased dielectric losses associated with this material.

The spacing between the two PCBs is maintained by 3D-printed spacers that can be snapped into mounting holes provided on both circuit boards. This low-cost solution ensures a well-defined and mechanically robust separation between the PCBs while keeping manufacturing and assembly simple and fast.

\begin{figure}[htbp]
  \centering
  \includegraphics[width=0.79\linewidth]{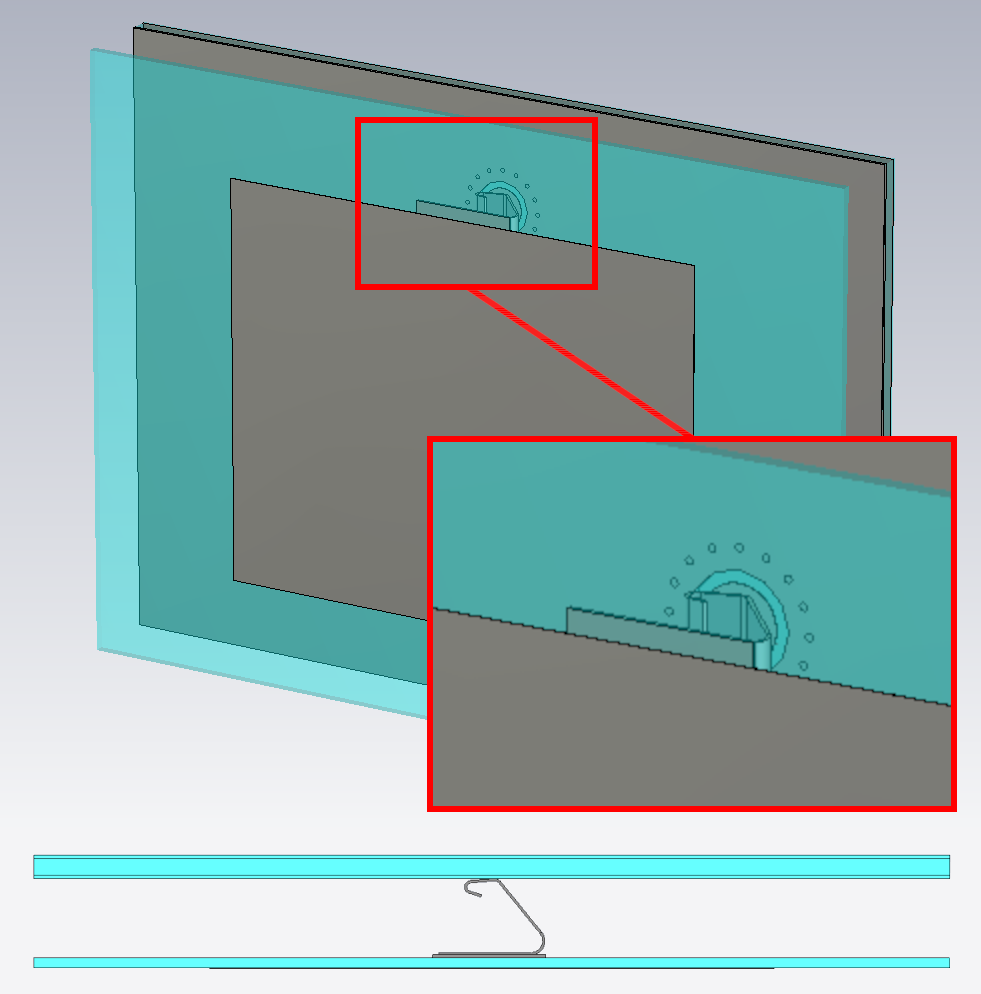}
  \caption{CST simulation model of the unit cell showing the feeding structure with an EMI spring contact. FR4 material is rendered in transparent blue.}
  \label{fig:unitCell}
\end{figure}

For patch antennas with the feed port located on the backside of the ground plane, a pin feed is commonly used. Typically, a wire pin is soldered through a hole in the patch on one end and connected to a through-hole pad on the other end. Since the goal is to enable RIS arrays with several hundred elements, manually soldering a wire feed for each unit cell is time-consuming and labor-intensive. In addition, pin-fed patch antennas exhibit a high input reactance due to the inductance of the feed pin, which becomes particularly problematic when the spacing between the patch and the ground plane is large. To address both issues, the wire pin is replaced by a spring contact that is typically used in electromagnetic interference (EMI) shielding applications. This spring contact is a machine-solderable component that can be automatically placed onto the patch PCB. To compensate for the inductance introduced by the spring contact, it is not connected directly to the patch. Instead, it is soldered on a pad located on the opposite side of the patch PCB, creating a capacitive coupling between the patch and the contact pad. The resulting combination forms a series resonant circuit, reducing the reactance of the feeding structure. Fig.~\ref{fig:unitCell} shows a 3D model of the unit cell as simulated in CST Studio Suite, with a side-view at the bottom.

The final dimensions of the optimized unit cell are a width of $w_\mathrm{uc} = \SI{60.0}{\milli\meter}$, a length of $l_\mathrm{uc} = \SI{45.0}{\milli\meter}$, a spacing between the patch and the ground plane of $d = \SI{5.0}{\milli\meter}$, a PCB substrate thickness of $t = \SI{0.8}{\milli\meter}$, a patch width of $w_\mathrm{p} = \SI{37.0}{\milli\meter}$, and a patch length of $l_\mathrm{p} = \SI{30.0}{\milli\meter}$. This unit cell is used for both versions with \SI{1}{\bit} and \SI{3}{\bit} phase resolution. The material parameters of the FR4 substrate were characterized by measurement, resulting in a relative permittivity of $\varepsilon_\mathrm{r} = 4.9$ and a dissipation factor of $\tan\delta = 0.025$.

\subsection{Phase Switching Circuits}
From CST, the simulated S-parameters of the unit cell were exported as a Touchstone file. Considering the Floquet port as Port 1 and the microstrip line port as Port 2, the overall reflection coefficient of the RIS, $\Gamma_\mathrm{RIS_\mathrm{i}}$, for a known load reflection coefficient $\Gamma_\mathrm{\varphi_\mathrm{i}}$ provided by the switching network, is given by
\begin{equation}\label{eq:S11fromSandGamma}
\Gamma_\mathrm{RIS_\mathrm{i}} = S_{11} + \frac{S_{21}S_{12}\Gamma_\mathrm{\varphi_\mathrm{i}}}{1 - S_{22}\Gamma_\mathrm{\varphi_\mathrm{i}}}.
\end{equation}
To accurately determine the reflection coefficient $\Gamma_\mathrm{\varphi_\mathrm{i}}$ of the phase-switching circuits, the two RF switches were characterized using a through-reflect-line (TRL) calibration kit on the same substrate and layer stack-up as used for the final RIS tile. Based on the measured S-parameters of the switches, the circuitry was modeled in Advanced Design System (ADS) to optimize the phase response of the RIS, $\mathrm{arg}\left (\Gamma_\mathrm{RIS} \right )$.

\subsubsection{1-Bit RF Switch}
For the 1-bit switch, a SKYA21024 single-pole double-throw (SPDT) RF switch from Skyworks is used. By leaving one port open and connecting the other port to ground, switching between $\Gamma_{\varphi_0} = \SI{0}{\degree}$ and $\Gamma_{\varphi_1} = \SI{180}{\degree}$ is achieved.

\subsubsection{3-Bit RF Switch}
\begin{figure}[htbp]
  \centering
  \begin{circuitikz}[scale=0.9, transform shape]
    % Transmission lines

% TLines

% State 0
\draw[draw=black, fill=white] (0,2.8) circle (0.05);
\draw[draw=black] (0.05,2.8) -- (0.2,2.8);
\draw[draw=black, fill=white, thick] (0.2,2.7) rectangle (0.8,2.9);
\node[thick, align=left, anchor=west] at (1.75,2.8) {\scriptsize state 0};
\node[thick, align=center, anchor=center] at (0.9,3.1) {\scriptsize microstrip lines };
% State 1
\draw[draw=black, fill=white] (0,2.4) circle (0.05);
\draw[draw=black] (0.05,2.4) -- (0.2,2.4);
\draw[draw=black, fill=white, thick] (0.2,2.3) rectangle (1,2.5);
\node[thick, align=left, anchor=west] at (1.75,2.4) {\scriptsize state 1};
% State 2
\draw[draw=black, fill=white] (0,2) circle (0.05);
\draw[draw=black] (0.05,2) -- (0.2,2);
\draw[draw=black, fill=white, thick] (0.2,1.9) rectangle (1.2,2.1);
\node[thick, align=left, anchor=west] at (1.75,2) {\scriptsize state 2};
% State 3
\draw[draw=black, fill=white] (0,1.6) circle (0.05);
\draw[draw=black] (0.05,1.6) -- (0.2,1.6);
\draw[draw=black, fill=white, thick] (0.2,1.5) rectangle (1.4,1.7);
\node[thick, align=left, anchor=west] at (1.75,1.6) {\scriptsize state 3};
% State 4
\draw[draw=black, fill=white] (0,1.2) circle (0.05);
\draw[draw=black] (0.05,1.2) -- (0.2,1.2);
\draw[draw=black, fill=white, thick] (0.2,1.1) rectangle (0.8,1.3);
\draw[draw=black] (0.8,1.2) -- (1,1.2) -- (1,1);
\draw[draw=black, thick] (0.9,1) -- (1.1,1);
\node[thick, align=left, anchor=west] at (1.75,1.2) {\scriptsize state 4};
% State 5
\draw[draw=black, fill=white] (0,0.8) circle (0.05);
\draw[draw=black] (0.05,0.8) -- (0.2,0.8);
\draw[draw=black, fill=white, thick] (0.2,0.7) rectangle (1,0.9);
\draw[draw=black] (1,0.8) -- (1.2,0.8) -- (1.2,0.6);
\draw[draw=black, thick] (1.1,0.6) -- (1.3,0.6);
\node[thick, align=left, anchor=west] at (1.75,0.8) {\scriptsize state 5};
% State 6
\draw[draw=black, fill=white] (0,0.4) circle (0.05);
\draw[draw=black] (0.05,0.4) -- (0.2,0.4);
\draw[draw=black, fill=white, thick] (0.2,0.3) rectangle (1.2,0.5);
\draw[draw=black] (1.2,0.4) -- (1.4,0.4) -- (1.4,0.2);
\draw[draw=black, thick] (1.3,0.2) -- (1.5,0.2);
\node[thick, align=left, anchor=west] at (1.75,0.4) {\scriptsize state 6};
% State 7
\draw[draw=black, fill=white] (0,0) circle (0.05);
\draw[draw=black] (0.05,0) -- (0.2,0);
\draw[draw=black, fill=white, thick] (0.2,-0.1) rectangle (1.4,0.1);
\draw[draw=black] (1.4,0) -- (1.6,0) -- (1.6,-0.2);
\draw[draw=black, thick] (1.5,-0.2) -- (1.7,-0.2);
\node[thick, align=left, anchor=west] at (1.75,0) {\scriptsize state 7};

% Switch
\draw[draw=black, thick] (-0.965,1.435) -- (-0.035,2.765);
\draw[draw=black, fill=white] (-1,1.4) circle (0.05);
\draw[draw=black] (-1.05,1.4) -- (-1.25,1.4);
\draw[draw=black, fill=white] (-1.3,1.4) circle (0.05);

% Phi labels
\draw[-stealth] (-0.7,2.2) to [bend left=45] (-0.4,1.9);
\node[thick, align=center, anchor=center] at (-0.55,2.8) {$\Gamma_\mathrm{\varphi_\mathrm{0}}$};
\draw[-stealth] (-0.3,2.8) -- (-0.1,2.8);

\node[thick, align=center, anchor=center] at (-0.55,0.9) {$\vdots$};

\node[thick, align=center, anchor=center] at (-0.55,0) {$\Gamma_\mathrm{\varphi_\mathrm{7}}$};
\draw[-stealth] (-0.3,0) -- (-0.1,0);

% S-Parameter
\draw[draw=black] (-1.35,1.4) -- (-1.6,1.4);
\draw[draw=black, fill=white, thick] (-2.2,1.1) rectangle (-1.6,1.7);
\draw[draw=black] (-2.2,1.4) -- (-2.45,1.4);
\draw[draw=black, fill=white] (-2.5,1.4) circle (0.05);

\node[thick, align=center, anchor=center] at (-1.9,1.4) {[S]};
\node[thick, align=center, anchor=center] at (-1.9,0.9) {\scriptsize antenna};
\node[thick, align=center, anchor=center] at (-2.5,1.6) {\scriptsize 1};
\node[thick, align=center, anchor=center] at (-1.3,1.6) {\scriptsize 2};

% GammaRIS
\draw[-stealth] (-3.15,1.4) -- (-2.65,1.4);
\node[thick, align=center, anchor=center] at (-3.65,1.4) {$\Gamma_\mathrm{RIS_\mathrm{i}}$};
  \end{circuitikz}
  \caption{Simplified schematic of the \SI{3}{\bit} unit cell.}
  \label{fig:schematic}
\end{figure}
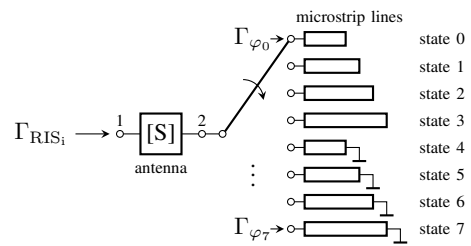

The 3-bit switch used is a SKY13418-485LF single-pole eight-throw (SP8T) RF switch, also from Skyworks Inc. To achieve phase switching in steps of $\SI{45}{\degree}$, microstrip transmission lines of different lengths are connected to the eight output ports of the switch, where half of the lines are left open and the other half are terminated to ground. This configuration keeps the line lengths short, thereby reducing losses and minimizing dispersion effects for a flat phase response over frequency. The individual line lengths were optimized in ADS to achieve reflection coefficients of $\Gamma_{\varphi_\mathrm{i}} = i \times \SI{45}{\degree}$ for the switching states $i = 0, 1, \ldots, 7$. Fig.~\ref{fig:schematic} shows the simplified schematic of the 3-bit unit cell.

\section{Realization} \label{sec:realization}
For each of the developed unit cells with \SI{1}{\bit} and \SI{3}{\bit} phase resolution, RIS tiles containing $4 \times 4$ elements were designed and fabricated. Each tile integrates a microcontroller on its back side for configuration of the RIS patterns, allowing multiple tiles to be concatenated into large-scale arrays that cover multiple square meters of RIS area.
To facilitate experiments with the RIS tiles, each unit cell is equipped with a red-green-blue (RGB) LED that indicates its currently configured switching state. The color assignments corresponding to the phase states are listed in Table~\ref{tab:colorAssignment} and are used consistently in the measurement result plots throughout this paper.

\begin{table}[htbp]
    \renewcommand{\arraystretch}{1.3}
    \caption{Color assignment.}
    \label{tab:colorAssignment}
    \centering
    \begin{tabular}{|c|c||c|c|}
        \hline
        Color & Phase & Color & Phase \\
        \hline
        \hline
        black & \SI{0}{\degree} & cyan & \SI{45}{\degree} \\
        \hline
        red & \SI{90}{\degree} & magenta & \SI{135}{\degree}\\
        \hline
        green & \SI{180}{\degree} & yellow & \SI{225}{\degree} \\
        \hline
        blue & \SI{270}{\degree} & white & \SI{315}{\degree}\\
        \hline
    \end{tabular}
\end{table}

\begin{figure}[htbp]
  \centering
  \includegraphics[width=0.9\linewidth]{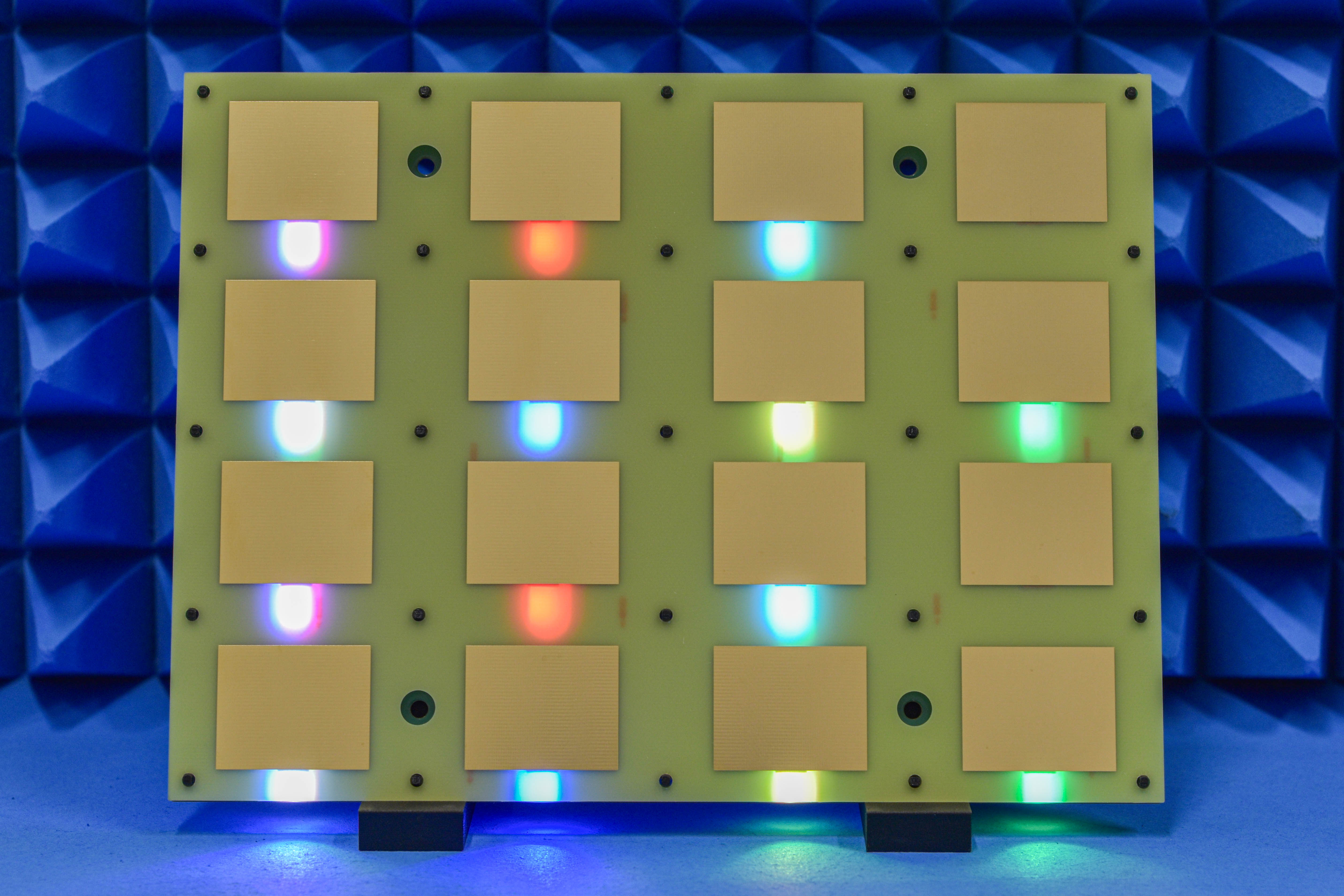}
  \caption{RIS tile with 16 unit cells and \SI{3}{\bit} phase resolution.}
  \label{fig:risTile}
\end{figure}

A single $4 \times 4$ RIS tile of the 3-bit version is shown in Fig.~\ref{fig:risTile}, where the configured states of each unit cell are indicated by the LEDs visible through the thin FR4 substrate layer. The power consumption of the phase switching circuitry is \SI{2.2}{\milli\watt} per tile, excluding the power required by the microcontroller and LEDs.

\begin{figure}[htbp]
  \centering
  \includegraphics[width=0.9\linewidth]{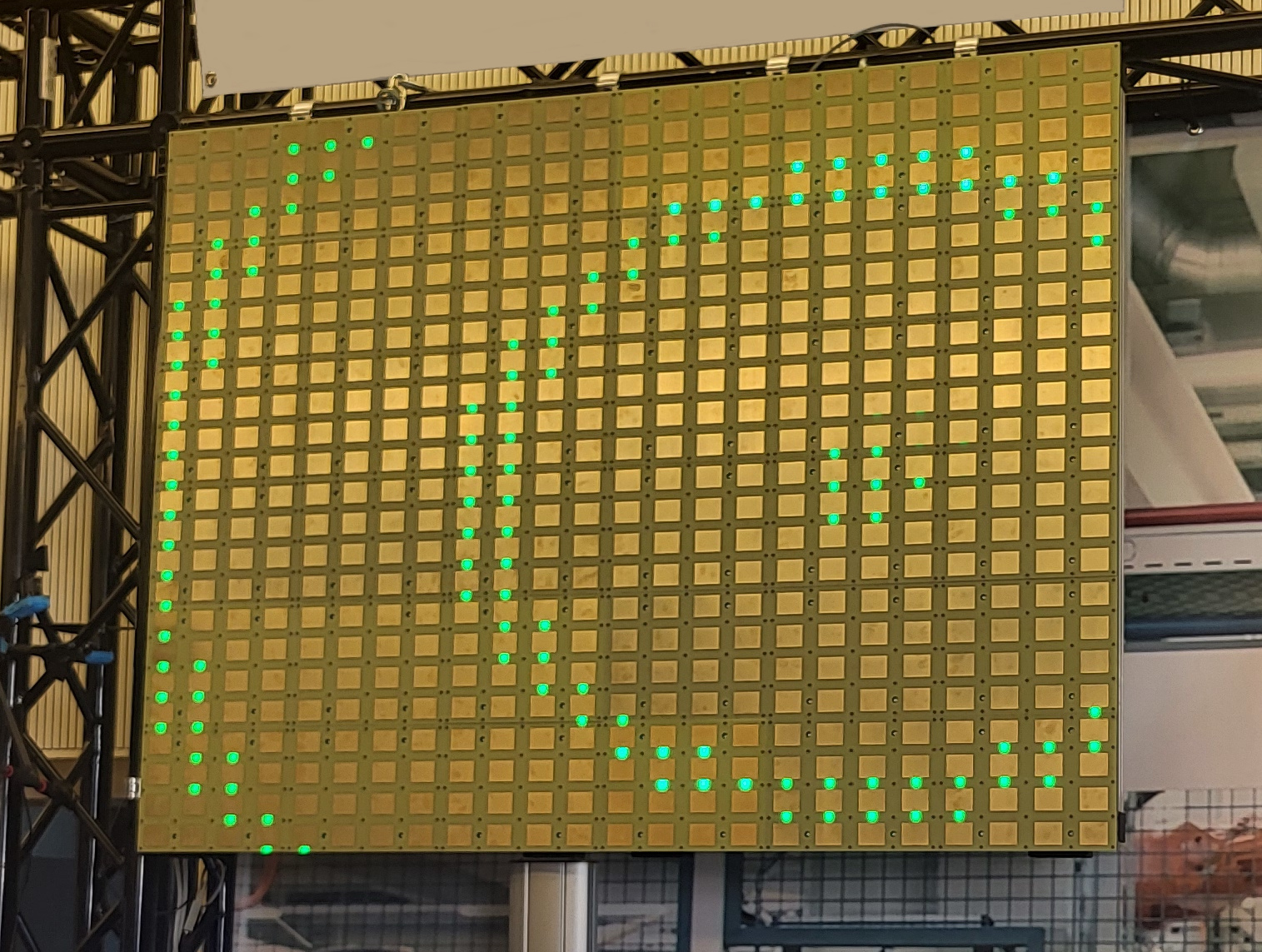}
  \caption{RIS with 576 unit cells and \SI{1}{\bit} phase resolution, size \SI{1.56}{\meter\squared}.}
  \label{fig:risWall}
\end{figure}

For the 1-bit version, an array consisting of $6 \times 6 = 36$ tiles, corresponding to a total of $24 \times 24 = 576$ elements, was assembled, resulting in an RIS area of approximately \SI{1.56}{\meter\squared}. The large-scale array is shown in Fig.~\ref{fig:risWall}, where it is mounted on an aluminum truss. The power consumption of the phase switching circuitry of a single 1-bit tile is \SI{0.20}{\milli\watt}, resulting in a total power consumption of \SI{7.2}{\milli\watt} for the complete array of 36 tiles, not accounting for microcontrollers or LEDs.

\section{Measurements} \label{sec:measurements}
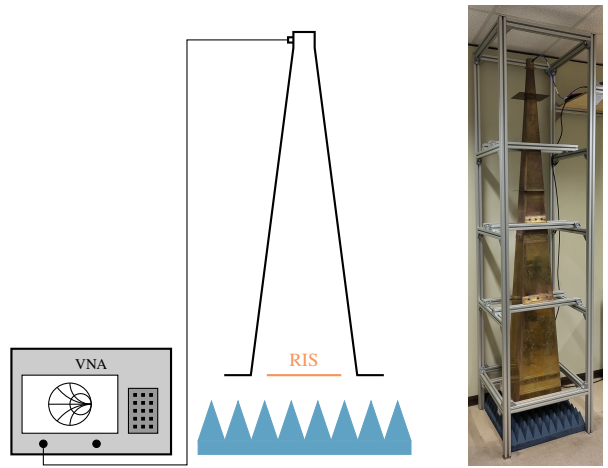
\begin{figure}[htbp]
  \centering
  \begin{circuitikz}[scale=0.7, transform shape]
    % VNA
\draw[draw=black, fill=gray!40, thick] (0,0) rectangle (3,2);
\node[thick, align=center, anchor=center] at (1.5,1.7) {\footnotesize VNA};
\draw[draw=black, fill=white] (0.2, 0.4) rectangle (2,1.5);
\draw[draw=black, fill=black] (0.6,0.2) circle (0.07);
\draw[draw=black, fill=black] (1.6,0.2) circle (0.07);
\draw[draw=black, fill=gray!80] (2.2,0.4) rectangle (2.75,1.25);
\draw[draw=black, fill=black] (2.3, 0.5) rectangle ++(0.05, 0.05);
\draw[draw=black, fill=black] (2.45, 0.5) rectangle ++(0.05, 0.05);
\draw[draw=black, fill=black] (2.6, 0.5) rectangle ++(0.05, 0.05);
\draw[draw=black, fill=black] (2.3, 0.65) rectangle ++(0.05, 0.05);
\draw[draw=black, fill=black] (2.45, 0.65) rectangle ++(0.05, 0.05);
\draw[draw=black, fill=black] (2.6, 0.65) rectangle ++(0.05, 0.05);
\draw[draw=black, fill=black] (2.3, 0.8) rectangle ++(0.05, 0.05);
\draw[draw=black, fill=black] (2.45, 0.8) rectangle ++(0.05, 0.05);
\draw[draw=black, fill=black] (2.6, 0.8) rectangle ++(0.05, 0.05);
\draw[draw=black, fill=black] (2.3, 0.95) rectangle ++(0.05, 0.05);
\draw[draw=black, fill=black] (2.45, 0.95) rectangle ++(0.05, 0.05);
\draw[draw=black, fill=black] (2.6, 0.95) rectangle ++(0.05, 0.05);
\draw[draw=black, fill=black] (2.3, 1.1) rectangle ++(0.05, 0.05);
\draw[draw=black, fill=black] (2.45, 1.1) rectangle ++(0.05, 0.05);
\draw[draw=black, fill=black] (2.6, 1.1) rectangle ++(0.05, 0.05);
\draw[draw=black] (1.1,0.95) circle (0.4);
\draw[draw=black] (1.3,0.95) circle (0.2);
\draw[draw=black] (1.5,0.95) arc (270:180:0.4);
\draw[draw=black] (1.1,0.55) arc (180:90:0.4);
\draw[draw=black] (0.7,0.95) -- (1.5,0.95);
\draw[draw=black] (1.5,0.95) arc (-90:-143:0.8);
\draw[draw=black] (1.5,0.95) arc (90:143:0.8);

% Absorber
\draw [draw=MidnightBlue!50, fill=MidnightBlue!50] (3.51cm,0.0cm) rectangle ++(4.0cm,0.25cm);
\foreach \i in {0,1,2,...,7}{
    \node[isosceles triangle, isosceles triangle stretches, rotate=90, draw=MidnightBlue!50, fill=MidnightBlue!50, minimum width=0.5cm, minimum height=0.75cm, anchor=left corner] (T)at (\i*0.5cm+3.5cm,0.25cm){};
}

% Horn antenna
\draw[draw=black, thick] (4.0,1.5) -- (4.5,1.5) -- (5.3,7.7) -- (5.3,8) -- (5.7,8) -- (5.7,7.7) -- (6.5,1.5) -- (7.0,1.5);
\draw[draw=black, thick] (5.3,7.8) -- (5.2,7.8) -- (5.2,7.9) -- (5.3,7.9);

% RIS
\draw[draw=Peach, thick] (4.8,1.5) -- (6.2,1.5);
\draw (5.5,1.8) node[color=Peach, align=center, anchor=center] {RIS};

% Lines
\draw[draw=black] (0.6,0.2) -- (0.6,-0.2) -- (3.3,-0.2) -- (3.3,7.85) -- (5.2,7.85);
  \end{circuitikz}
  \hspace{0.5cm}
  \includegraphics[width=0.2\linewidth]{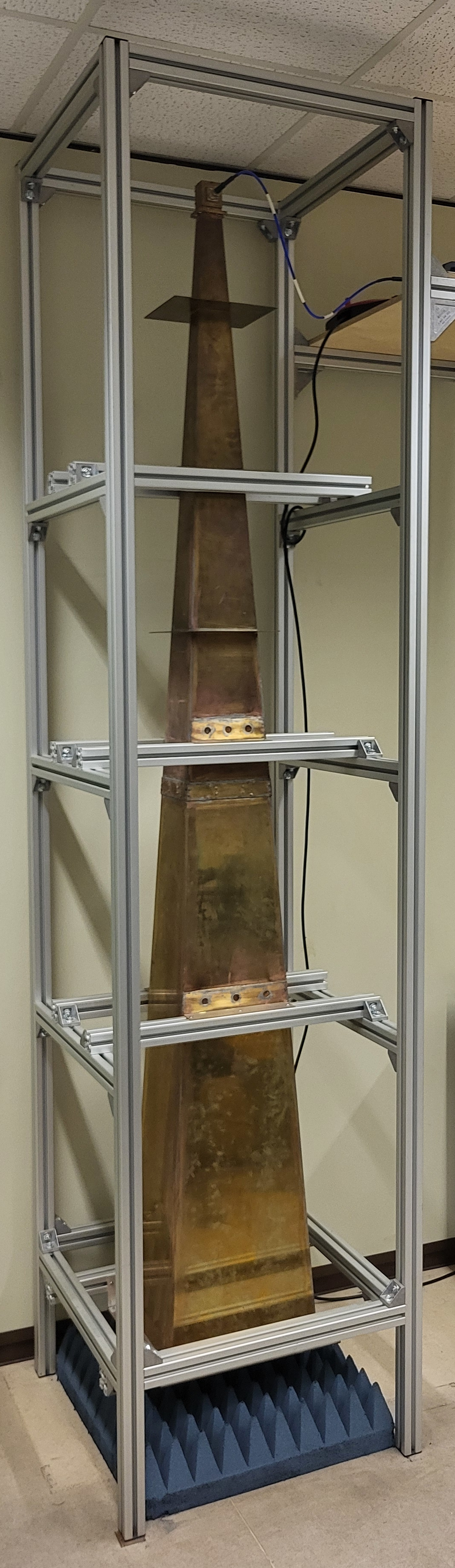}
  \caption{Horn antenna measurement setup.}
  \label{fig:measurementSetup}
\end{figure}

The reflection coefficients of the two RIS versions were experimentally characterized using the horn-antenna setup shown in Fig.~\ref{fig:measurementSetup}, where a single RIS tile was positioned in the antenna aperture above RF absorbers. The return loss of the horn antenna was measured with a Streamline P9373A vector network analyzer (VNA) from Keysight to extract the reflection coefficient of the RIS tile, $\Gamma_\mathrm{RIS}$, employing a time-gating procedure to exclude reflections occurring at the coax-to-waveguide transition of the horn. The measured values were then normalized to a metallic reference plate of the same size as the RIS tile, providing an absolute magnitude and phase reference. In this way, the mono-static reflection coefficient for a wave incident and reflected normal to the RIS surface could be obtained. During the measurement, all 16 unit cells of the RIS tile under test were configured to the same switching state being evaluated.

\begin{figure*}[htbp]
\centering
\subfloat[Magnitude response, \SI{1}{\bit} version.]{\includegraphics[width=1.0\columnwidth]{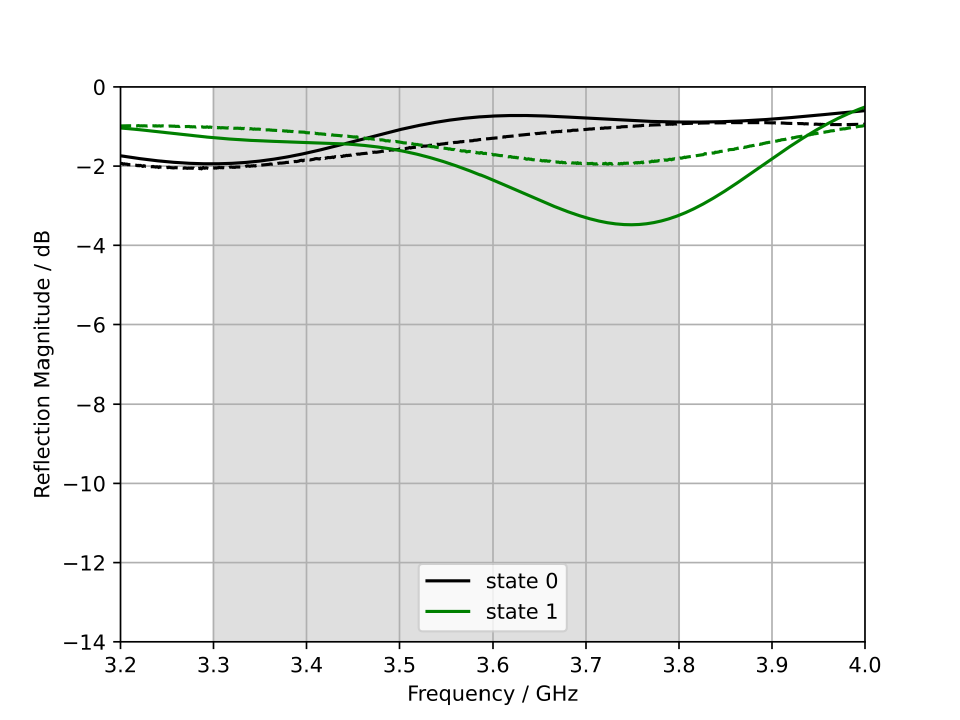}
\label{fig:mag1bit}}
\hfil
\subfloat[Magnitude response, \SI{3}{\bit} version.]{\includegraphics[width=1.0\columnwidth]{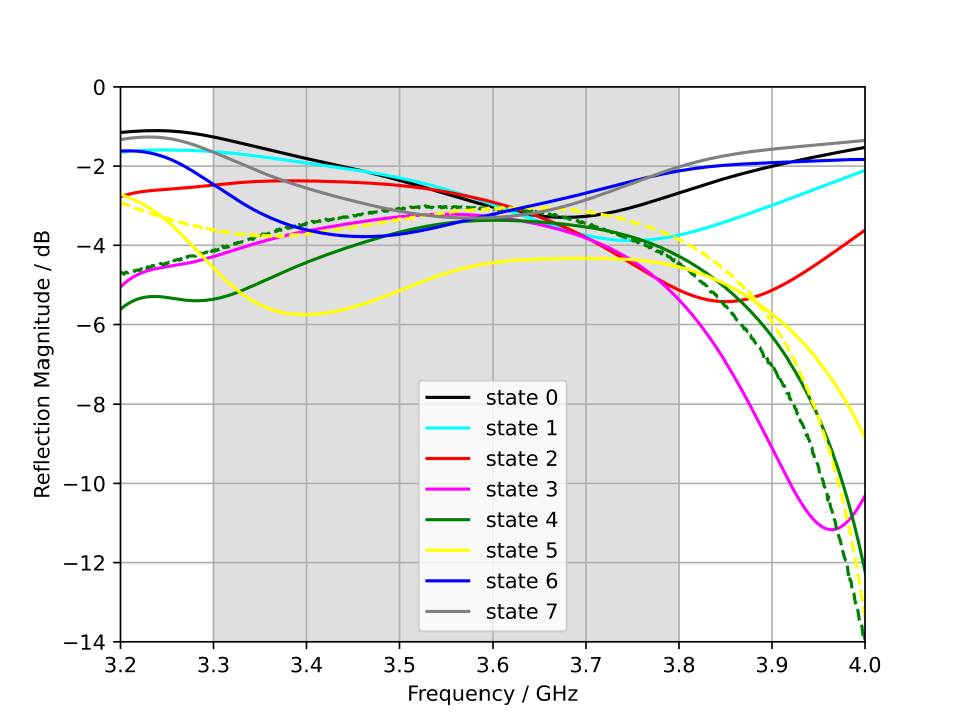}
\label{fig:mag3bit}}\\
\subfloat[Phase response normalized to state 0, \SI{1}{\bit} version.]{\includegraphics[width=1.0\columnwidth]{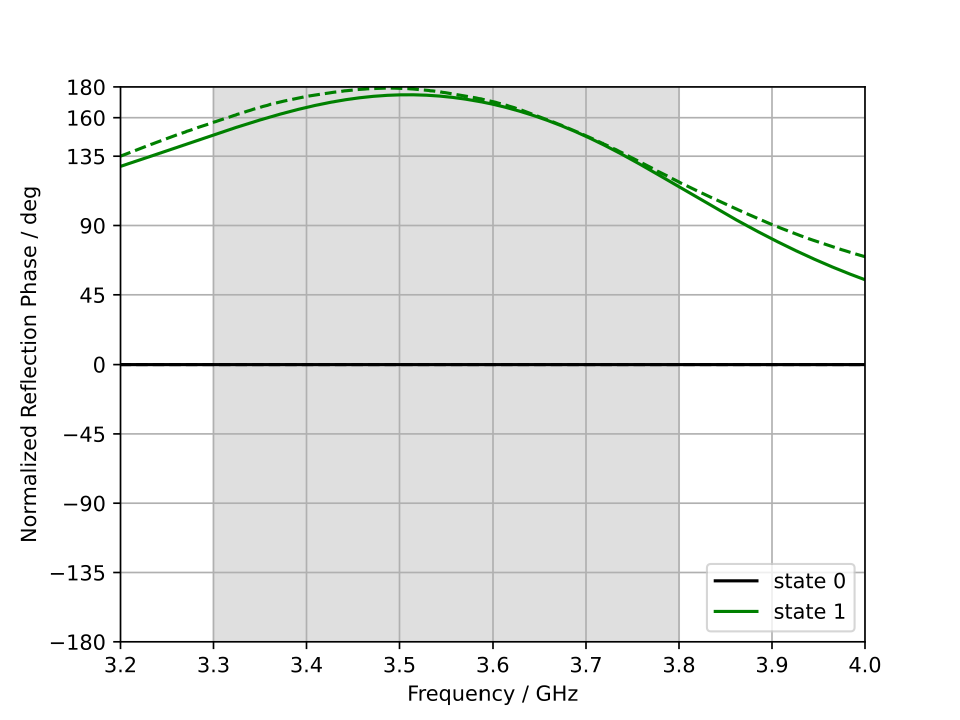}
\label{fig:phase1bit}}
\hfil
\subfloat[Phase response normalized to state 0, \SI{3}{\bit} version.]{\includegraphics[width=1.0\columnwidth]{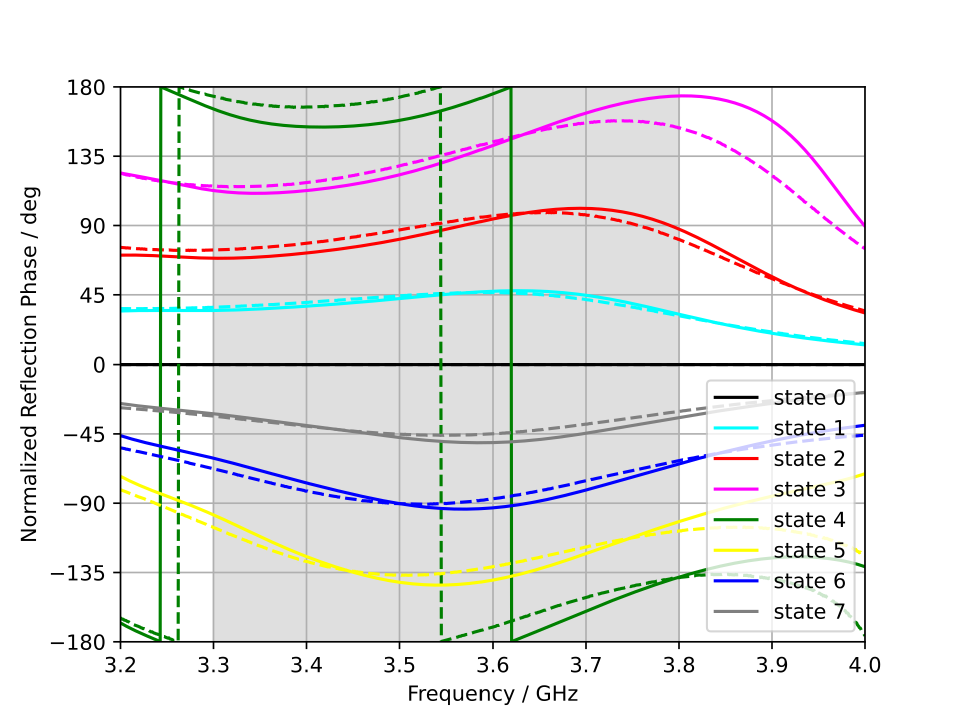}
\label{fig:phase3bit}}
\caption{Magnitude and phase responses of the \SI{1}{\bit} and \SI{3}{\bit} RIS tiles, n78 frequency band highlighted in gray, measurement solid, simulation dashed.}
\label{fig:measurementResults}
\end{figure*}

The measured magnitude and phase responses of the 1-bit RIS tile are shown in Figs.~\ref{fig:mag1bit} and \ref{fig:phase1bit}, with the n78 frequency band ($3.3-\SI{3.8}{\giga\hertz}$ in Germany) highlighted in gray. Within this band, the lowest magnitude occurs in state 1 and is \SI{-3.5}{\decibel}. The phase difference between states 0 and 1 remains at least \SI{100}{\degree} across the entire \SI{500}{\mega\hertz} n78 band.

The measurement results of the 3-bit RIS tile are presented in Figs.~\ref{fig:mag3bit} and \ref{fig:phase3bit}. At the designed center frequency of \SI{3.6}{\giga\hertz}, the magnitude response is approximately \SI{-3.2}{\decibel} for all eight switching states, except for state~5, which exhibits a value of about \SI{-4.3}{\decibel}. Across the entire n78 band, the lowest magnitude is observed for state~5, reaching \SI{-5.75}{\decibel}. For improved visual clarity, only the simulated magnitude responses for state~4 and state~5 are presented, where state~4 exhibits the best agreement between measurement and simulation, while state~5 shows the largest discrepancy. For the phase, all measured responses agree well with the simulated curves over the full operating band. The phase response is optimal at \SI{3.60}{\giga\hertz}, where the deviation of the measured normalized phase from the target switching-state phase is minimal.

\section{Comparison with other RIS designs} \label{sec:comparison}
For comparison of the proposed RIS with existing designs reported in the literature, the bandwidth definition introduced in~\cite{Pereira} is adopted. In~\cite{Pereira}, the standard deviation $\sigma$ is defined as
\begin{equation}
\sigma =
\sqrt{\frac{\sum_{i=0}^{2^{n}-1} \left(\Delta\varphi_i\right)^3}{12 \times \SI{360}{\degree}}},
\end{equation}
where $n$ denotes the number of bits, resulting in $2^{n}$ switching states, and $\Delta\varphi_i$ represents the phase difference between two adjacent switching states, as illustrated in Figs.~6(c) and~6(d). Based on the standard deviation, the effective number of bits $N_\mathrm{bit}$ is derived as
\begin{equation}
N_\mathrm{bit} = \log_2 \left( \frac{\SI{360}{\degree}}{\sqrt{12}\,\sigma} \right).
\end{equation}
As a boundary condition on the phase performance, a minimum effective bit number of $N_\mathrm{bit} = \SI{1.7}{\bit}$ is used for the 2-bit design in \cite{Pereira}. Consequently, for a 1-bit, 2-bit, and 3-bit RIS, the corresponding minimum effective resolutions are derived to be $N_{\mathrm{bit,1bit}} = \SI{0.7}{\bit}$, $N_{\mathrm{bit,2bit}} = \SI{1.7}{\bit}$, and $N_{\mathrm{bit,3bit}} = \SI{2.7}{\bit}$. This results in maximum allowable standard deviations of $\sigma_{\mathrm{1bit}} \le \SI{65}{\degree}$, $\sigma_{\mathrm{2bit}} \le \SI{32.5}{\degree}$, and $\sigma_{\mathrm{3bit}} \le \SI{16.25}{\degree}$.

\subsection{1-bit Version}
Using the above mentioned bandwidth limit of $\sigma_\mathrm{1bit} \leq \SI{65}{\degree}$, the usable bandwidth of the 1-bit version is determined to be greater than \SI{500}{\mega\hertz}, covering the whole n78 frequency band. For comparison, the patch-antenna-based RIS design reported in \cite{Rains}, operating at a center frequency of \SI{3.5}{\giga\hertz} and fabricated on a low-loss RF substrate with a dielectric constant of $\varepsilon_\mathrm{r} = 2.2$, a dissipation factor of $\tan\delta = 0.001$, and a thickness of \SI{2}{\milli\meter}, achieves a bandwidth of \SI{255}{\mega\hertz} with a maximum insertion loss of $\SI{4.2}{\decibel}$. In that design, phase switching is implemented using a varactor diode, and the unit cell size is $\SI{30.0}{\milli\meter} \times \SI{30.0}{\milli\meter}$.

\balance

\subsection{3-bit Version}
Three different RIS designs from the literature were considered for comparison of the 3-bit version. The first design, reported in \cite{Zhu}, employs a dipole structure backed by a ground plane as the antenna element. The PCB substrate has a dielectric constant of $\varepsilon_\mathrm{r} = 2.2$ and a dissipation factor of $\tan\delta = 0.001$. Two substrate layers are stacked, resulting in a total thickness of \SI{4}{\milli\meter} between the ground plane and the dipoles. The unit cell dimensions are $\SI{30.1}{\milli\meter} \times \SI{30.1}{\milli\meter}$. Phase switching is realized using two positive-intrinsic-negative (PIN) diodes, providing a phase resolution of \SI{2}{\bit}, corresponding to a step size of \SI{90}{\degree}. Accordingly, the bandwidth is defined as the frequency range where the standard deviation remains below $\SI{32.5}{\degree}$, yielding a bandwidth of approximately \SI{109}{\mega\hertz}.

In \cite{Dey}, an RIS design based on a ground plane-backed split convoluted square ring frequency selective surface (FSS) is proposed, with two different versions presented. The first version employs a substrate with a dielectric constant of $\varepsilon_\mathrm{r} = 3.2$, a dissipation factor of $\tan\delta = 0.003$, and a thickness of \SI{1.58}{\milli\meter}, combined with an air gap of \SI{1.0}{\milli\meter} between the substrate and the ground plane. The unit cell size is $\SI{8.0}{\milli\meter} \times \SI{8.0}{\milli\meter}$, significantly smaller than the RIS presented in this paper ($\SI{60}{\milli\meter} \times \SI{45}{\milli\meter}$) due to the different operating principle. Phase switching is realized using varactor diodes. This first version provides \SI{2}{\bit} phase resolution at a center frequency of $\SI{3.8}{\giga\hertz}$, resulting in a usable bandwidth of approximately \SI{190}{\mega\hertz}.
The second version is designed for operation at a center frequency of \SI{4.2}{\giga\hertz} with a phase resolution of \SI{3}{\bit}. It employs a substrate with a dielectric constant of $\varepsilon_\mathrm{r} = 2.2$, a dissipation factor of $\tan\delta = 0.0009$, and a thickness of \SI{1.58}{\milli\meter}, without an air gap. The unit cell size is identical to that of the first version. This second variant provides a usable bandwidth of approximately \SI{85}{\mega\hertz}.

When using the above mentioned bandwidth constraints from \cite{Pereira}, one has to consider that the allowable phase deviation is reduced with an increasing phase resolution. Thus, for a valid comparison of the 3-bit RIS presented in this work with 2-bit RIS versions from the literature, the bandwidth of a virtual 2-bit version of the presented RIS is calculated. For this, only the phase states 0, 2, 4 and 6 are considered for computation of the standard deviation. This results in a 2-bit bandwidth of more than \SI{500}{\mega\hertz}, covering the complete span of the n78 band.

\subsection{Section Summary}

As summarized in Table~\ref{tab:compare3bit}, the proposed RIS provides a significantly higher bandwidth compared to recent designs in the literature, for all versions with \SI{1}{\bit}, virtual \SI{2}{\bit}, and \SI{3}{\bit} phase resolution.

\balance

\begin{table}[htbp]
    \renewcommand{\arraystretch}{1.3}
    \caption{Bandwidth comparison with the literature.}
    \label{tab:compare3bit}
    \centering
    \begin{tabular}{|c|c|c|c|c|}
        \hline
        Unit Cell & Frequency & Resolution & Bandwidth \\
        \hline
        \hline
        Proposed design & \SI{3.6}{\giga\hertz} & \multirow{2}{*}{\SI{1}{\bit}} & \SI{>500}{\mega\hertz} \\
        \cline{1-2} \cline{4-4}
        \cite{Rains} & \SI{3.5}{\giga\hertz} & & \SI{255}{\mega\hertz} \\
        \hline\hline
        Proposed design & \SI{3.6}{\giga\hertz} & \multirow{3}{*}{\SI{2}{\bit}} & \SI{>500}{\mega\hertz} \\
        \cline{1-2} \cline{4-4}
        \cite{Zhu} & \SI{3.5}{\giga\hertz} & & \SI{109}{\mega\hertz} \\
        \cline{1-2} \cline{4-4}
        \cite{Dey} & \SI{3.8}{\giga\hertz} & & \SI{190}{\mega\hertz} \\
        \hline
        \hline
        Proposed design & \SI{3.6}{\giga\hertz} & \multirow{2}{*}{\SI{3}{\bit}} & \SI{440}{\mega\hertz} \\
        \cline{1-2} \cline{4-4}
        \cite{Dey} & \SI{4.2}{\giga\hertz} & & \SI{85}{\mega\hertz} \\
        \hline
    \end{tabular}
\end{table}

\section{Conclusion} \label{sec:conclusion}
In this work, a scalable reconfigurable intelligent surface for \SI{3.6}{\giga\hertz} 5G/6G applications has been presented. The proposed design supports both 1-bit and 3-bit phase resolution, offering broadband operation across the n78 frequency band. The combination of a patch-antenna-based unit cell with a novel spring-contact feeding structure enables efficient assembly and reduces manufacturing complexity for large-scale arrays. Experimental results demonstrate that the RIS achieves wideband phase tunability with low insertion loss, surpassing the usable bandwidth of comparable designs reported in the literature. The modular tile-based architecture, low power consumption, and clear visual feedback through RGB LEDs make the design well-suited for experimental studies of smart radio environments, beam steering, and sensing applications in next-generation wireless networks.

\section*{Acknowledgment}
This work was funded by the German Federal Ministry of Research, Technology and Space (BMFTR) in the framework of project Reflect6G under grant numbers 16KIS2229 and 16KIS2228K.

\end{document}